\documentclass[11pt,onecolumn]{IEEEtran}

\ifCLASSINFOpdf
  \usepackage[pdftex]{graphicx}
 \else
  \usepackage[dvips]{graphicx}
\fi

\usepackage[cmex10]{amsmath}
\usepackage{amssymb}

\usepackage{array}

\usepackage{amssymb}
\usepackage{enumitem}
\usepackage{verbatim}
\usepackage{bbold}

\newtheorem{thm}{Theorem}

\newtheorem{lem}{Lemma}
\newtheorem{prop}{Proposition}

\renewcommand{\mod}{\mathrm{mod}}

\newcommand{\beq}{\begin{equation}}
\newcommand{\eeq}{\end{equation}}

\newcommand{\bv}{\mathbf{v}}
\newcommand{\bx}{\mathbf{x}}
\newcommand{\by}{\mathbf{y}}
\newcommand{\bard}{\bar{d}}
\newcommand{\bzero}{\mathbf{0}}

\makeatletter

\newcommand{\Rmnum}[1]{\expandafter\@slowromancap\romannumeral #1@}
\makeatother
\usepackage{mathtools}
\DeclarePairedDelimiter\ceil{\lceil}{\rceil}

\begin{document}
%
\title{Fundamental Limits on Data Acquisition: Trade-offs between Sample Complexity and Query Difficulty}

\author{Hye Won Chung$^*$,~Ji Oon Lee,~and Alfred O. Hero
\thanks{Hye Won Chung$^*$ (hwchung@kaist.ac.kr) is with the School of Electrical Engineering at KAIST in South Korea. Ji Oon Lee (jioon.lee@kaist.edu) is with the Department of Mathematical Sciences at KAIST in South Korea. Alfred O. Hero (hero@eecs.umich.edu) is with the Department of EECS at the University of Michigan. This work was partially supported by National Research Foundation of Korea under grant number 2017R1E1A1A01076340, and by United States Army Research Office under grant W911NF-15-1-0479. 
}}

\maketitle


\begin{abstract}
We consider query-based data acquisition and the corresponding information recovery problem, where the goal is to recover $k$ binary variables (information bits) from parity measurements of those variables.
The queries and the corresponding parity measurements are designed using the encoding rule of Fountain codes. By using Fountain codes, we can design potentially limitless number of queries, and corresponding parity measurements, and guarantee that the original $k$ information bits can be recovered with high probability from any sufficiently large set of measurements of size $n$. In the query design, the average number of information bits that is associated with one parity measurement is called query difficulty ($\bar{d}$) and the minimum number of measurements required to recover the $k$ information bits for a fixed $\bar{d}$ is called sample complexity ($n$). We analyze the fundamental trade-offs between the query difficulty and the sample complexity, and show that the sample complexity of $n=c\max\{k,(k\log k)/\bar{d}\}$ for some constant $c>0$ is necessary and sufficient to recover $k$ information bits with high probability as $k\to\infty$. 
\end{abstract}



%




\IEEEpeerreviewmaketitle

\section{Introduction}

Query-based data acquisition arises in diverse applications including: crowdsourcing~\cite{karger2014budget, bernstein2011crowds}; active learning~\cite{mackay1992information,settles2010active}; experimental design~\cite{lindley1956measure,fedorov1972theory}; and community recovery or clustering in graphs~\cite{abbe2015community,hajek2017information}. 
In these applications, query-based data acquisition can be modeled as a 20 questions problem~\cite{chung2017unequal} between an oracle (or oracles) and a player where the oracle knows the values of information bits that the player aims to recover, while the player designs queries to the oracle and receives answers from the oracle.
In this paper, we consider query-based data acquisition with the goal of recovering the values of $k$ variables $(x_1,\dots, x_k)$ (information bits). 
When we assume that the variables and the answers (measurements) are binary, we can consider a parity check sum as a type of measurement, which corresponds to exclusive or (XOR or modulo 2 sum) of some subset of the information bits. 
Querying parity symbols of the information bits generalizes the 20 questions model of~\cite{chung2017unequal,tsiligkaridis2014collaborative}. This generalization is the focus of this paper and has a wide range of applications, in particular to crowdsourcing systems~\cite{karger2014budget}.

Consider a crowdsourcing system consisting of a number of workers and a particular task that they are expected to work on. 
Assume that the task is to classify a collection of $k$ images into two exclusive groups, e.g., whether or not an image is suitable for children.
A worker (oracle) in the system is given a query about some subset of the images and asked to provide a binary answer regarding those images. 
Assume that the worker can skip the query if the worker is unsure of the answer. The probability that one worker skips a query is unknown at the stage of query design and it can be different for each of the workers in the system, depending on their abilities or efforts. 
Therefore, from the query designer's point of view, it is natural to assume that only a random subset of the designed queries will be answered by the workers in the crowdsourcing system. The query designer's objective is to design queries such that when the received number of answers exceeds some threshold, regardless of which subset of the answers was collected, the original $k$ binary bits can be recovered with high probability. 

In this paper we show that Fountain codes~\cite{mackay2005fountain}  are naturally suited to this crowdsourcing query design problem. 
Fountain codes are a type of forward error correcting codes suitable for binary erasure channels (BEC) with unknown erasure probabilities. This type of codes has been the subject of much research for reliable internet packet transmissions when the packets transmitted from the source are randomly lost before they arrive at the destination. 
For $k$ input symbols $(x_1,x_2,\dots, x_k)$, Fountain codes produce a potentially limitless number of parity measurements, which are also called output symbols. By using well-designed Fountain codes, one can guarantee that, given any set of output symbols of size $k(1+\delta)$ with small overhead $\delta>0$, the input symbols can be recovered with high probability.  Examples of Fountain codes are LT-codes~\cite{luby2002lt} and Raptor codes~\cite{shokrollahi2006raptor}. 
By using these Fountain codes, we can design potentially limitless number of queries (and corresponding parity measurements) with the desired properties suitable for the crowdsourcing example.

However, the Fountain code framework must be extended in order to account for the worker's limited capacity to answer difficult queries.  
The \emph{query difficulty} is defined as the average number of input symbols required to compute a single parity measurement. The query difficulty is related to the encoding complexity for one-stage encoding, but it is different from the encoding complexity when the encoding is done in multiple stages. The query difficulty represents the number of input symbols  on average the worker must know to calculate one parity measurement. 
Depending on the query difficulty, the number of answers (parity measurements) required to recover $k$ input symbols may vary greatly. We call the minimum number of measurements required to recover $k$ input symbols the \emph{sample complexity}. The sample complexity is a function of the query difficulty as well as the number $k$ of input symbols to be recovered.

Let us consider two extreme cases. First, consider the case when the query difficulty is equal to 1. More specifically, we assume that each query asks the value of only one variable in $(x_1,x_2,\dots, x_k)$ at a time.
Since it is not known which of the queries would be answered by the workers at the stage of query design, a set of queries is designed by uniformly and randomly picking one variable in $(x_1,x_2,\dots, x_k)$ at a time.
For such querying scenarios, with randomly selected $n$ measurements, in order to recover the $k$ information bits with error probability less than $1/k^u$ for some constant $u>0$, the required number $n$ of queries scales as $k\log k$. On the other hand, when each query is designed to generate a parity measurement of randomly selected $k/2$ bits at a time, the required number $n$ of measurements is only $k+c\log k$ for some constant $c>0$~\cite{shokrollahi2006raptor}. Therefore, in these two extreme cases, we can observe that when the query difficulty is equal to 1, the sample complexity scales as $k\log k$, whereas for the query difficulty of order $k$ the sample complexity scales as $k$. Then the question is how the sample complexity scales as the query difficulty increases from $1$ to $\Theta(k)$.


In this paper, we aim to analyze the fundamental trade-offs between the sample complexity and the query difficulty in recovering $k$ information bits.
There have been papers that have analyzed such trade-offs when it is assumed that the parity measurements involve only a fixed number $1\leq d\leq k$ of input symbols. 
In~\cite{chen2016information}, the case of pairwise measurements ($d=2$) was considered and in~\cite{ahn2016community}, a general integer $1\leq d\leq k$ was considered. 
Note that there are a total of $k\choose d$ possible parity measurements for a fixed $d$. In both papers, it was assumed that each measurement is independently observed with probability $p_{\sf obs}$.
It was shown that the number $n$ of measurements to recover $k$ input symbols with high probability scales as $n=p_{\sf obs}{k\choose d}=c\max\{k,({k\log k})/{d}\}$ for some constant $c>0$. 

In this paper, we generalize the work in~\cite{ahn2016community} by not fixing the number $d$ but instead allowing that the number $d$ follows a distribution $(\Omega_0,\Omega_1,\dots, \Omega_k)$ where $\Omega_d$ denotes the probability that the value $d$ is chosen, where $\sum_{d=0}^k \Omega_d=1$. We consider the average query difficulty $\bar{d}=\sum_{d=0}^k d\,\Omega_d$ and analyze the sample complexity $n$ as a function of the query difficulty $\bar{d}$ and the number $k$ of input symbols.
By assuming that $d$ follows the prescribed distribution, we can generate potentially limitless parity measurements by using the encoding rule employed by Fountain codes; guaranteeing that for any set of fixed number of measurements it is possible to recover the $k$ input symbols with high probability.
This framework is thus more suitable for the situations where the parity measurements are erased with arbitrary (unknown) probabilities and thus it is required to have the ability to generate potentially limitless number of queries and corresponding parity measurements. 
For the previous frameworks in~\cite{chen2016information,ahn2016community}, the maximum number of parity measurements is restricted to $k\choose d$ for a fixed $d$.

Our main contribution in this paper is to specify the fundamental trade-offs between the sample complexity ($n$) and the query difficulty ($\bard$) in this generalized measurement model. 
We show that the sample complexity $n$ necessary and sufficient to recover $k$ input symbols with high probability scales as
\beq\label{eqn:mainres}
n=c\cdot \max\left\{k,(k\log k)/{\bar{d}}\right\}
\eeq
for some constant $c>0$. 
Note that for $\bar{d}= O(\log k)$, the sample complexity $n$ is inversely proportional to the query difficulty $\bar{d}$. In particular, when query difficulty $\bar{d}=\Theta(1)$, the sample complexity scales as $k\log k$, whereas when $\bar{d}=\Theta(\log k)$, the sample complexity scales as $k$.

The rest of this paper is organized as follows. In Section~\ref{sec:model}, we explain the encoding rule of Fountain codes (how to generate potentially limitless number of parity measurements) and state the main problem of this paper. In Section~\ref{sec:main}, we provide the main results, showing the fundamental trade-offs between the sample complexity and the query difficulty. 
In Section~\ref{sec:proof}, we prove the main theorem. More technical details for the proof are presented in Appendices. In Section~\ref{sec:sim}, some simulation results are provided, which further support our theoretical results.
In Section~\ref{sec:con}, we provide conclusions and discuss possible future research directions. 

\subsection{Notations}
We use the notation $\oplus$ for XOR of binary variables, i.e., for $a,b\in\{0,1\}$, $a\oplus b=0$ iff $a=b$ and $a\oplus b=1$ iff $a\neq b$.
We denote by $e_j$ the $k$-dimensional unit vector with its $j$-th element equal to 1.
For a vector $\bx$, $\|\bx\|_1$ denotes the number of $1$'s in the vector $\bx$.
For vectors $\bx$ and $\by$, the inner product between $\bx$ and $\by$ is denoted by $\bx \cdot \by$.
For two integers $\alpha$ and $\beta$, we use the notation $\alpha\equiv \beta$ to indicate that $\mod(\alpha,2)=\mod(\beta,2)$.
For two vectors $\bx=(x_1,x_2,\dots, x_k)$ and $\by=(y_1,y_2,\dots, y_k)$, when we write $\bx\equiv \by$, it means that $\mod(x_i,2)=\mod(y_i,2)$ for all $i\in\{1,2,\dots,k\}$.
We use the $O(\cdot)$ and $\Theta(\cdot)$ notations to describe the asymptotics of real sequences $\{a_k\}$ and $\{b_k\}$: $a_k=O(b_k)$ implies that $a_k\leq M b_k$ for some positive real number $M$ for all $k\geq k_0$; $a_k=\Theta(b_k)$ implies that $a_k\leq M b_k$ and $a_k \geq M' b_k$ for some positive real numbers $M$ and $M'$ for all $k\geq k_0'$.
The logarithmic function $\log$ is with base $e$.

\section{Model and Problem Statement}\label{sec:model}
Consider a $k$-dimensional binary random vector $\bx=(X_1, X_2,\dots, X_k)^T$, which is uniformly and randomly distributed over $\{0,1\}^k$. We call $X_1,X_2,\dots, X_k$ the input symbols. 
We aim to learn the values of $(X_1, X_2,\dots, X_k)$ by observing a total of $n$ parity measurements of different subsets of those $k$ bits. 
Consider $k$-dimensional binary vectors $\mathbf{v}_i=(v_{i1},v_{i2},\dots, v_{ik})$, $i=1,\dots,n$. The parity measurement associated with the vector $\mathbf{v}_i$ is defined by
\beq
\begin{aligned}
Y_i&=\mod\left( \sum_{j=1}^k v_{ij} X_j, 2\right)=v_{i1}X_1\oplus \dots\oplus v_{ik}X_k,
\end{aligned}
\eeq
for $i=1,\dots, n$. We call such parity measurements $(Y_1,Y_2,\dots, Y_n)$ the output symbols. 
Each $\mathbf{v}_i\in\{0,1\}^k$ determines which subset of $(X_1, X_2,\dots, X_k)$ is to be picked in calculating the $i$-th parity measurement.

The process of designing $\{\mathbf{v}_i\}$ is called query design or encoding. We use Fountain codes, also known as erasure rateless codes, for the encoding. 
Let $(\Omega_0,\Omega_1,\dots, \Omega_k)$ be a distribution on $\{0,1,\dots, k\}$ where $\Omega_d$ denotes the probability that the value $d$ is chosen and $\sum_{d=0}^k \Omega_d=1$.  
In the encoding of Fountain codes, each vector $\mathbf{v}_i$ is generated independently and randomly by first sampling a weight $d\in\{0,1,\dots,k\}$ from the distribution $(\Omega_0,\dots, \Omega_k)$ and then selecting a $k$-dimension vector of weight $d$ uniformly at random from all the vectors of $\{0,1\}^k$ with weight $d$. 

\begin{figure}[t]
\centerline{\includegraphics[scale=0.5]{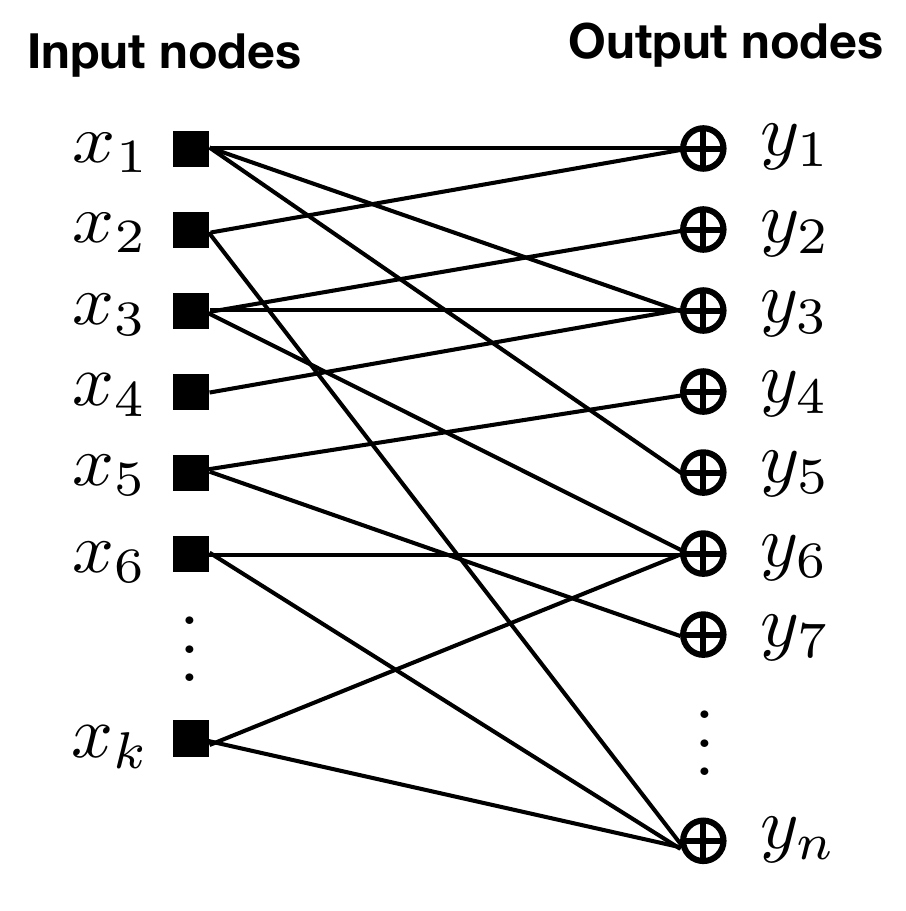}}
\caption{Bipartite graph between input nodes and output nodes.}
\label{fig:graph}
\end{figure}

Consider an arbitrary set of $n$ output symbols $(Y_1,\dots, Y_n)$ generated by the above encoding rule. The relationship between the $k$ input symbols and the $n$ output symbols can be depicted by a bipartite graph with $k$ input nodes on one side and $n$ output nodes on the other side as shown in Fig~\ref{fig:graph}.
Denote by $\bar{d}$ the average degree of the output nodes, 
\beq
\bar{d}=\sum_{d=1}^k d\cdot \Omega_d.
\eeq
This number $\bar{d}$ indicates the average number of input symbols involved in one parity measurement (output symbol) and is related to the difficulty in calculating one parity measurement. We call this number \emph{query difficulty}. 

The process of recovering the $k$ input symbols from the $n$ output symbols is called information recovery or decoding. 
Denote by $\hat{\mathbf{x}}(Y_1^k)$ the estimate of $\bx$ given $Y_1^n$ and define the probability of error as 
\beq\label{eqn:err_prob}
P_e^{(k)}=\min_{\hat{\bx}(\cdot)}\Pr(\hat{\bx}(Y_1^k)\neq \bx).
\eeq
With the proper choice of the distribution  $(\Omega_0,\dots, \Omega_k)$, the Fountain codes guarantee that $P_e^{(k)}\to 0$ as $k\to\infty$ with $n$ larger than some threshold.
The minimum number of $n$ required to guarantee $P_e^{(k)}\to 0$ as $k\to\infty$, minimized over all $(\Omega_0,\dots, \Omega_k)$ for a fixed $k$ and $\bard$, is called \emph{sample complexity}.
We aim to find the fundamental limits on $n$ to guarantee reliable information recovery of $k$ input symbols for a fixed query difficulty $\bar{d}$. 

\section{Main Results: Fundamental Trade-offs between Sample Complexity and Query Difficulty}\label{sec:main}

In this section, we state our main results that the sample complexity $n$ that is necessary and sufficient to make $P_e^{(k)}\to 0$ as $k\to\infty$ scales in terms of $k$ and $\bar{d}$ as $n=c\cdot\max\{k,({k\log k})/{\bar{d}}\}$ for some constant $c>0$ independent of $k$ and $\bar{d}$, when the parity measurements are generated by the encoding rule of Fountain codes as explained in Section~\ref{sec:model}. 

We first state the well-known lower bound on the sample complexity $n$ of Fountain codes presented in~\cite{shokrollahi2006raptor}.
\begin{prop}\label{prop:suff}
{\it
To reliably recover $k$ input symbols with $P_e^{(k)}\leq 1/k^u$ for some constant $u>0$ from parity measurements generated by Fountain codes, it is necessary that
\beq\label{eqn:lower}
n\geq c_l\max\left\{k,\frac{k\log k}{\bar{d}}\right\}
\eeq
for some constant $c_l>0$.
}
\end{prop}
\begin{IEEEproof}
Showing the first condition $n\geq c_l k$ is straightforward. Each output symbol $Y_i=\mod\left(\sum_{j=1}^k v_{ij} X_j,2\right)$ represents a linear equation of $k$ unknown input symbols $(X_1,X_2,\dots, X_k)$. Since there are $k$ unknowns, it is necessary to have at least $n=k$ linear equations to solve this linear system reliably. 

The second condition $n\geq (c_l {k\log k})/{\bard}$ is from a property of random graphs. In the bipartite graph between input nodes and output nodes, we say that an input node is isolated if it is not connected to any of the output nodes. We analyze the probability that an input node is isolated when the edges are designed by the encoding rule of Fountain codes. The error probability $P_e^{(k)}$ is bounded below by the probability that an input node is isolated, since when an input node is isolated the decoding error happens. 

Consider an output node with degree $d$. The probability that an input node is not connected to this output node of degree $d$ equals $1-d/k$. Since an output node has degree $d$ with probability $\Omega_d$, the probability that an input node is not connected to an output node equals
\beq
\sum_{d=0}^k \Omega_d (1-d/k)=1-\bar{d}/k.
\eeq
Since there are $n$ output nodes and these output nodes are sampled independently, the probability that an input node is isolated (not connected to any of those output nodes) equals
\beq
\left(1-\frac{\bar{d}}{k}\right)^{n}.
\eeq
By the mean value theorem, we can show that $(1-\bar{d}/{k})^{n}\geq e^{-\alpha/(1-\alpha/n)}$ where $\alpha=n\bar{d}/k$.
Since the decoding error probability $P_e^{(k)}$ is lower bounded by the probability that an input node is isolated, to satisfy $P_e^{(k)}\leq 1/k^u$ for some constant $u>0$, it is necessary that $e^{-\alpha/(1-\alpha/n)}\leq 1/k^u$, which is equivalent to
\beq
\begin{split}
\alpha&\geq \log k\cdot\frac{u}{1+(u\log k)/n}\\
&\geq \log k \cdot \frac{u}{1+(u\log k)/k}\\
&\geq \log k\cdot  \frac{u}{1+(u\log 3)/3}\\
&\geq c_l\log k
\end{split}
\eeq
for some constant $c_l>0$.
By plugging in $\alpha=n\bar{d}/k$, we get
\beq
n\bar{d}\geq c_l k\log k.
\eeq
\end{IEEEproof}

The main contribution of this paper is showing that the bound in~\eqref{eqn:lower} is indeed achievable (up to constant scaling) by properly designed Fountain codes for any $\bar{d}$ from $\Theta(1)$ to $\Theta(\log k)$.
We provide a particular output degree distribution $(\Omega_0,\Omega_1,\dots, \Omega_k)$ for which we can control the query difficulty $\bar{d}$ from $\Theta(1)$ to $\Theta(\log k)$ and show that it is possible to reliably recover $k$ information bits with sample complexity obeying
\beq\label{eqn:eq}
n= c_u \max\left\{k,\frac{k\log k}{\bar{d}}\right\}
\eeq
for some constant $c_u>0$. Therefore, by combining~\eqref{eqn:eq} with~\eqref{eqn:lower}, we conclude that $n=c\cdot\max\{k,({k\log k})/{\bar{d}}\}$ is necessary and sufficient for reliable recovery of $k$ information bits when $\bar{d}$ is the average query difficulty. 

Suppose that the law of $\Omega_d$ is given by an ideal Soliton distribution
\beq\label{eqn:modSol}
\begin{aligned}
\Omega_d = 
	\begin{cases}
	\frac{1}{D} & \text{ if } d=1 \\
	\frac{1}{d(d-1)} & \text{ if } 2\leq d \leq D \\
	0 & \text{ if } d > D\text{ or }d=0,
	\end{cases}
\end{aligned}
\eeq
for some $D\in\{2,3,\dots,k\}$. Here, for simplicity, we assume that $k\geq 3$.
Note that the query difficulty scales as $\log D$ since
$$
\log (D+1) < \bar d = \frac{1}{D} + \sum_{d=2}^{D} \frac{1}{d-1} = \sum_{d=1}^D \frac{1}{d} < \log D + 1.
$$
Therefore, as $D$ increases from $2$ to $k$, the query difficulty $\bard$ scales from $\log 3$ to $\log k$.

\begin{thm}\label{thm:main}
{\it
For the Soliton distribution~\eqref{eqn:modSol} with $D\in\{2,3,\dots,k\}$, the $k$ input symbols can be reliably recovered, i.e., $P_e^{(k)}\to0$ as $k\to\infty$, with sample complexity
\beq\label{eqn:sam_up}
n=c_u \cdot \max\left\{k,\frac{k\log k}{\bar{d}}\right\}
\eeq
for some constant $c_u>0$. 
}
\end{thm}
The proof of Theorem~\ref{thm:main} will be presented in Section~\ref{sec:proof}.

Theorem~\ref{thm:main} states that for query difficulty $\bard = O(\log k)$, the sample complexity $n$ to reliably recover $k$ input symbols is inversely proportional to the query difficulty $\bard$. 
When the query difficulty does not increase in $k$, i.e., $\bard=\Theta(1)$, it is necessary and sufficient to have $n=\Theta(k\log k)$ to reliably recover the $k$ information bits.
In this regime, the ratio between $k$ and $n$ converges to 0 as $k\to\infty$.
On the other hand, when we increase the query difficulty to $\bard=\Theta(\log k)$, it is enough to have $n=\Theta(k)$ samples, which results in a positive limit of $k/n$ as $k\to\infty$.
When $\bard \gg \log k$, increasing the query difficulty no longer helps in reducing the sample complexity.

By using the Soliton distribution~\eqref{eqn:modSol} and the encoding rule of Fountain codes, we can design potentially limitless number of queries about $(x_1,x_2,\dots, x_k)$ and the corresponding parity measurements.  Theorem~\ref{thm:main} shows that with any set of measurements of size $n$ no larger than~\eqref{eqn:sam_up}, we can reliably recover the $k$ information bits as $k\to \infty$. 
Moreover, this sample size is optimal up to constants as shown by Proposition~\ref{prop:suff}.
Thus, our results provide the optimal query design strategy for reliable information recovery from an arbitrary set of parity measurements, optimal in terms of the sample complexity (up to constants) for a fixed query difficulty.

\section{Proof of Theorem~\ref{thm:main}}\label{sec:proof}
In this section, we prove Theorem~\ref{thm:main} by providing an upper bound on $P_e^{(k)}$ and showing that the sample complexity $n$ sufficient to make this upper bound converge to $0$ as $k\to\infty$ is equal to $c_u\cdot \max\{k,(k\log k)/\bard\}$ for some constant $c_u>0$.

Consider $P_e^{(k)}$ defined in~\eqref{eqn:err_prob}. The optimal decoding rule $\hat{\bx}(\cdot)$ that minimizes the probability of error is the maximum likelihood (ML) decoding for the uniformly distributed input symbols.
Assume that we collect $n$ parity measurements $(Y_1,\dots, Y_n)$ each of which equals $Y_i=\mod\left(\sum_{j=1}^n v_{ij} X_j,2\right)$. Consider a matrix $A$ whose $i$-th row is $\bv_i=(v_{i1},v_{i2},\dots, v_{ik})$, i.e.,
\beq
A:=[\bv_1;\bv_2;\dots; \bv_n].
\eeq
We call $A$ a sampling matrix. Given $(Y_1,\dots, Y_n)$ and the sampling matrix $A$, the ML decoding rule finds $\bx=(X_1,X_2,\dots, X_k)^T\in\{0,1\}^k$ such that
\beq
A\bx\equiv (Y_1,Y_2,\dots,Y_n)^T.
\eeq
If there is a unique solution $\bx\in\{0,1\}^k$ for this linear system, then it is claimed that $\hat{\bx}(Y_1^k)=\bx$. If there is more than one $\bx$ satisfying this linear system, then an error is declared. 
The probability of error is thus equal to
\beq
P_e^{(k)}=\sum_{\bx\in\{0,1\}^k} \frac{1}{2^k}\Pr(\exists \bx' \neq \bx \text{ such that }A\bx'\equiv A\bx).
\eeq
Due to symmetry, the probabilities $\Pr(\exists \bx' \neq \bx \text{ such that }A\bx'\equiv A\bx)$ are equal for every $\bx\in\{0,1\}^k$. Thus, we focus on the case where $\bx$ is the vector of all zeros and consider
\beq
P_e^{(k)}=\Pr(\exists \bx' \neq \bzero \text{ such that }A\bx'\equiv \bzero).
\eeq
By using the union bound, it can be shown that
\beq
\begin{aligned}
P_e^{(k)}
&\leq \sum_{\bx'\neq \bzero} \Pr(A\bx'\equiv \bzero)=\sum_{s=1}^k \sum_{\|\bx'\|_1=s}\Pr( A\bx'\equiv \bzero )\\
&=\sum_{s=1}^k { k \choose s} \Pr\left(A\left(\sum_{i=1}^s e_i\right) \equiv \bzero\right)
\end{aligned}
\eeq
where $e_i$ is the $i$-{th} standard unit vector. The last equality follows from the symmetry of the sampling matrix $A$. 
Since all the output samples are generated independently by the identically distributed $\bv_i$'s, each of which has weight $d$ with probability $\Omega_d$,
\beq
\begin{aligned}\label{eqn:P_e_m1}
&P_e^{(k)}\leq \sum_{s=1}^k { k \choose s}  \left(  \Pr\left( \bv_1\cdot \left(\sum_{i=1}^s e_i\right) \equiv 0\right) \right)^n\\
&=\sum_{s=1}^k \left({ k \choose s}\times\right.\\
&\quad\quad
\left.\left(\sum_{d=1}^k \Omega_d \Pr\left(\bv_1\cdot \left(\sum_{i=1}^s e_i\right) \equiv 0\Bigg| \|\bv_1\|_1=d\right) \right)^n\right).
\end{aligned}
\eeq
We next analyze 
\beq
\Pr\left(\bv_1\cdot \left(\sum_{i=1}^s e_i\right)\equiv0 \Bigg| \|\bv_1\|_1=d\right).
\eeq
Note that $\bv_1\cdot \left(\sum_{i=1}^s e_i\right)\equiv0$ if and only if there are even number of 1's in the first $s$ entries of $\bv_1$. This probability equals
\beq
\begin{split}\label{eqn:single}
&\Pr\left(\bv_1\cdot \left(\sum_{i=1}^s e_i\right)\equiv 0 \Bigg| \|\bv_1^T\|=d\right)=\frac{\sum_{\substack{i\leq d\\ i \text{ is even}}} {s \choose i}{k-s \choose d-i} }{{k \choose d}}.
\end{split}
\eeq

We next provide an upper bound on~\eqref{eqn:single}. Define
\beq
I_d = \sum_{\substack{i\leq d\\ i \text{ is even}}}{s\choose i}{k-s\choose d-i}.
\eeq
In the following lemma, we provide an upper bound on $I_d$ as a multiple of ${k\choose d}$. The proof of this lemma is based on that of the similar lemma provided in~\cite{ahn2016community}, where the upper bound on $I_d$ is stated depending on the regimes of $s$ for a fixed $d$. We provide an alternative version where the upper bound on $I_d$ depends on the regimes of $d$ for a fixed $s$. 

\begin{lem}\label{lem:comb}
{\it
Consider the case that $s \leq \frac{k}{2}$ (i.e., $s \leq k-s$). Define 
\beq\kappa(s) = \frac{k-s+1}{2s+1}.\eeq
\begin{enumerate}
\item For $d \leq \frac{k}{2}$ (or, $k-d \geq d$), when we define $\alpha = \frac{k-d+1}{d},$
	\beq
	I_d \leq \begin{cases}
	\left(1- \frac{2s}{5\alpha}\right){k\choose d},& \text{when } d < \kappa(s), \\
	\frac{4}{5}{k\choose d}, & \text{when } d \geq \kappa(s).
	\end{cases}
	\eeq
\item For $d > \frac{k}{2}$ (or, $k-d < d$), when we define  $\alpha' = \frac{d+1}{k-d}$,
	\beq
	I_d \leq \begin{cases}
	\left(1- \frac{2s}{5\alpha'}\right){k\choose d},& \text{when } d > k-\kappa(s), \\
	\frac{4}{5}{k\choose d}, & \text{when } d \leq k-\kappa(s).
	\end{cases}
	\eeq
\end{enumerate}
In the case $s > \frac{k}{2}$, we can obtain the bounds for $I_d$ simply by changing $s$ to $k-s$.
}
\end{lem}
\begin{IEEEproof}
Appendix~\ref{app:lem:comb}.
\end{IEEEproof}

By using Lemma~\ref{lem:comb} and~\eqref{eqn:single}, the upper bound on $P_e^{(k)}$ in~\eqref{eqn:P_e_m1} can be further bounded by
\beq \begin{split} \label{eqn:Sigma_s bound}
P_e^{(k)} &\leq 2\sum_{s \leq \frac{k}{2}} { k \choose s}\left(\sum_{d=1}^{\ceil*{\kappa(s)}-1} \left(1- \frac{2s}{5\alpha}\right) \Omega_d+ \sum_{d=\ceil*{\kappa(s)}}^{k-\ceil*{\kappa(s)}} \frac{4}{5}\Omega_d\right.\\
&\left. \qquad\qquad\qquad\quad + \sum_{d=k-\ceil*{\kappa(s)}+1}^{k} \left(1- \frac{2s}{5\alpha'}\right) \Omega_d \right)^n \\
&= 2\sum_{s \leq \frac{k}{2}} { k \choose s} (1-\Sigma_s)^n \leq 2\sum_{s \leq \frac{k}{2}} { k \choose s} e^{-n \Sigma_s},
\end{split} \eeq
where we let
\beq\label{eqn:sig_s}
\begin{split}
\Sigma_s &= \frac{1}{5} \sum_{d=\ceil*{\kappa(s)}}^{k-\ceil*{\kappa(s)}} \Omega_d + \\
&\frac{2s}{5} \left( \sum_{d=1}^{\ceil*{\kappa(s)}-1} \frac{d\,\Omega_d}{k-d+1} + \sum_{d=k-\ceil*{\kappa(s)}+1}^{k} \frac{(k-d)\Omega_d}{d+1} \right).
\end{split}
\eeq

Suppose that the law of $\Omega_d$ is given by a Soliton distribution provided in~\eqref{eqn:modSol}.
Here, for simplicity, we assume that $D\in\{2,3,\dots,k\}$ and $k\geq 3$.
For this Soliton distribution, we provide an upper bound on ${k \choose s}e^{-n\Sigma_s}$ in~\eqref{eqn:Sigma_s bound} for $s \leq \frac{k}{2}$ depending on the regime of $\ceil*{\kappa(s)}$ with conditions on the sample complexity $n$.
\begin{lem}\label{lem:kappas}
{\it
With the sample complexity
\beq\label{eqn:sam_suff1}
n\geq c_u\max\left\{k,\frac{k\log k}{\bar{d}}\right\}
\eeq
for some constant $c_u>0$, the term ${k \choose s}e^{-n\Sigma_s}$ is bounded above as follows. 
\begin{enumerate}
\item If $\ceil*{\kappa(s)} > D$,
\beq
\binom{k}{s} e^{-n\Sigma_s} <  k^{-s}.
\eeq
\item If $4 \leq \ceil*{\kappa(s)} \leq D$
\beq
\binom{k}{s} e^{-n\Sigma_s}  \leq
	\begin{cases}
	k^{-s} & \text{ if } s \leq \sqrt k \,, \\
	2^{-2\sqrt k} &\text{ if } \sqrt k < s \leq k/2 \,.
	\end{cases}
\eeq
\item If $\ceil*{\kappa(s)} \leq 3$,
\beq
\binom{k}{s} e^{-n\Sigma_s} \leq 2^k e^{-k}.
\eeq
\end{enumerate}
}
\end{lem}
\begin{IEEEproof}
Appendix~\ref{app:lem:kappas}.
\end{IEEEproof}
We remark that the case 2) does not happen when $D \in \{ 2, 3\}$. 

From Lemma~\ref{lem:kappas}, when the sample complexity $n$ satisfies~\eqref{eqn:sam_suff1} we can further bound $P_e^{(k)}$ in~\eqref{eqn:Sigma_s bound} by
\beq
\begin{aligned}
P_e^{(k)}&\leq 2\sum_{s \leq \frac{k}{2}} \left(k^{-s}+2^{-2\sqrt{k}}+2^k e^{-k}\right)\\
&\leq c'\left(\frac{1}{k}+k2^{-2\sqrt{k}}+k2^ke^{-k}\right)
\end{aligned}
\eeq
for some constant $c'>0$. Note that this upper bound converges to 0 as $k\to\infty$.

\section{Simulations}\label{sec:sim}
\begin{figure}[t]
\centerline{\includegraphics[scale=0.45]{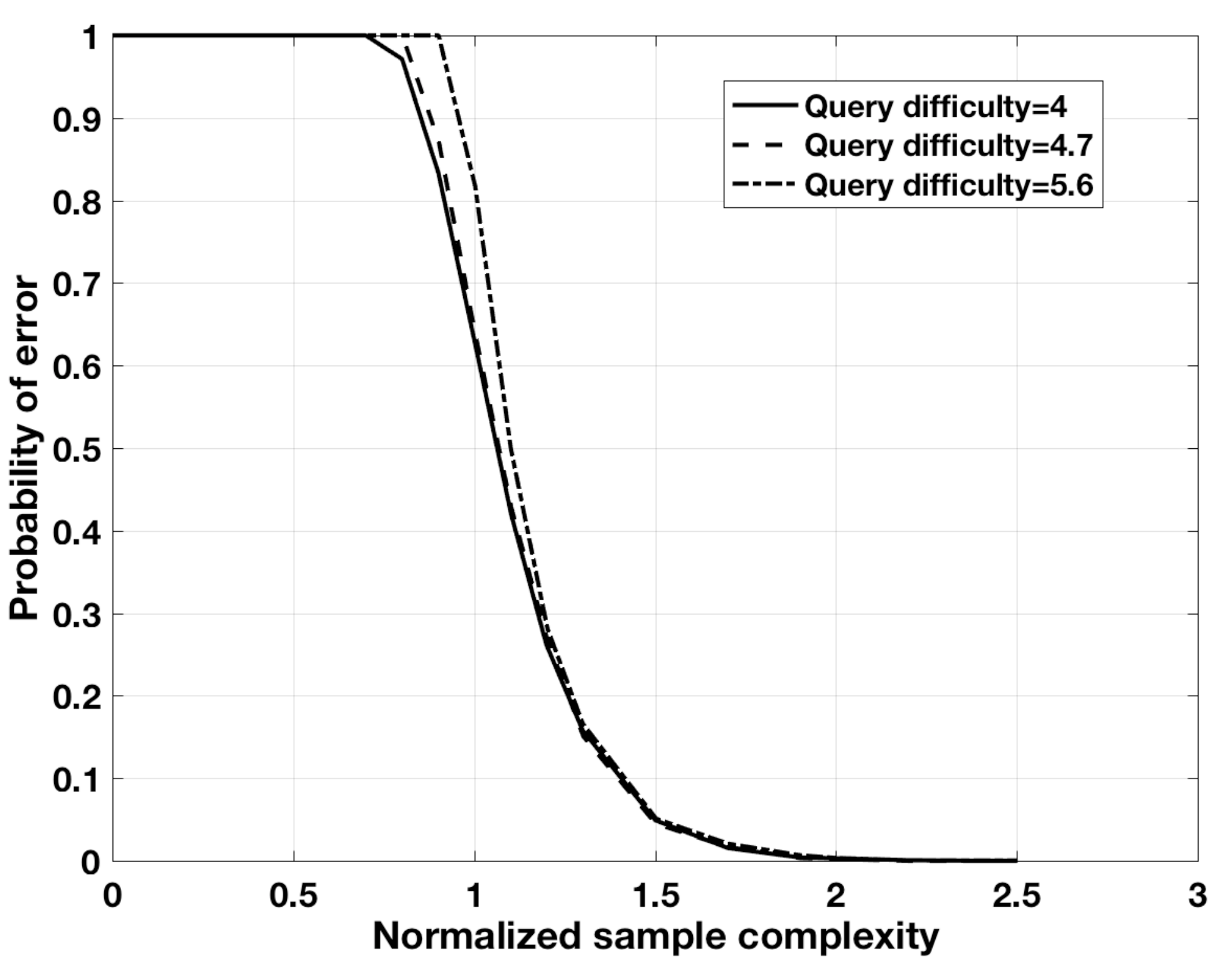}}
\caption{Monte Carlo simulation (5000 runs) of the probability of error $P_e^{(k)}$ with $k=300$ for three different $\bard$'s (the query difficulties). The sample complexity is normalized by $(k\log k)/\bard$. 
We can observe the phase transition for $P_e^{(k)}$ around the normalized sample complexity equal to 1 for all the three query difficulties considered.}
\label{fig:PeMC}
\end{figure}

\begin{figure}[t]
\centerline{\includegraphics[scale=0.45]{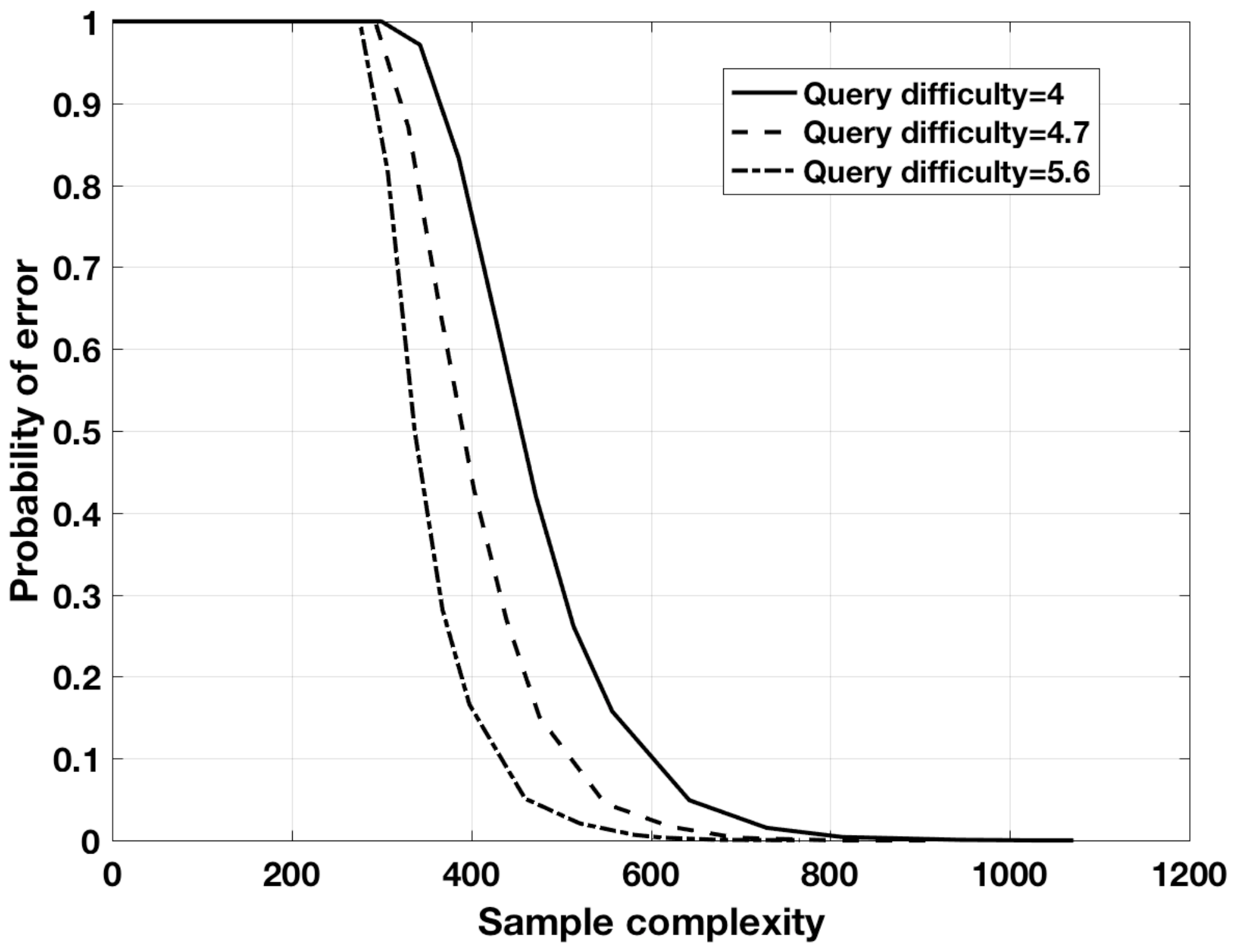}}
\caption{Same simulation conditions as in Fig~\ref{fig:PeMC} except that the horizontal axis is the un-normalized sample complexity. As the query difficulty increases, the sample complexity to make $P_e^{(k)}$ close to 0 decreases. This illustrates the trade-offs between the query difficulty and the sample complexity. }
\label{fig:PeMCnotnorm}
\end{figure}

In this section, we provide empirical performance analysis for the probability of error in the recovery of $k$ information bits, as a function of the sample complexity and query difficulty.
In Fig.~\ref{fig:PeMC}, we provide Monte Carlo simulation results for the probability of error $P_e^{(k)}$, defined in~\eqref{eqn:err_prob}, where the number $k$ of information bits to recover is fixed as $k=300$.
We plot $P_e^{(k)}$ in terms of the normalized sample complexity, normalized by $(k\log k)/\bard$ where $\bard$ is the query difficulty. We run the simulations for three different query difficulties, $\bard=$4, 4.7, 5.6.
The parity measurements (output symbols) are designed by first sampling $d$ (the number of input symbols required to compute a single parity measurement) from the Soliton distribution~\eqref{eqn:modSol} and then generating the measurements by the encoding rule of Fountain codes. 

Observe the phase transition of $P_e^{(k)}$ around the normalized sample complexity equal to 1. In Theorem~\ref{thm:main}, we stated that with sample complexity of $c_u\cdot \max\{k,(k\log k)/\bard\}$ for some constant $c_u>0$, we can guarantee $P_e^{(k)}\to 0$ as $k\to\infty$. The simulation results show that $c_u\approx 1$ is sufficient to produce a dramatic decrease of $P_e^{(k)}$. 
Since the phase transition occurs in the vicinity of normalized sample complexity equal to 1, the figure demonstrates the trade-offs between the query difficulty and the sample complexity. Specifically, the required number of parity measurements to reliably recover $k$ information bits is inversely proportional to the query difficulty when $\bard = O(\log k)$. 
Note that for the Soliton distribution~\eqref{eqn:modSol}, the query difficulty is $O(\log k)$, and thus $\max\{k,(k\log k)/\bard\}=\Theta( (k\log k)/\bard)$. 
In Fig~\ref{fig:PeMCnotnorm}, we show the same simulation with un-normalized sample complexity indexing the horizontal axis. From this plot, we can observe that as the query difficulty increases, the required number of samples to make $P_e^{(k)}$ close to 0 decreases.

\section{Conclusions}\label{sec:con}

In this paper, we analyzed the fundamental trade-offs between query difficulty $\bard$ and sample complexity $n$ in a query-based data acquisition system associated with  a crowdsourcing task with workers who may be non-responsive to certain queries (channel erasures).
We considered the information recovery of $k$ binary variables $(x_1,x_2,\dots, x_k)$ from parity measurements of subsets of these variables. 
We used a query design based on the encoding rules of Fountain codes, with which we can design potentially limitless numbers of queries. We showed that the proposed query design policy  guarantees that the original $k$ information bits can be recovered with high probability from any set of measurements of size $n$ larger than some threshold.
We obtained necessary and sufficient conditions on sample complexity $n\geq c\cdot \max\{k, (k\log k)/\bard\}$. 

There are several interesting future research directions related to this work. 
One of such directions includes analyzing trade-offs between query difficulty and sample complexity for partial information recovery problems.
In this paper, we considered exact information recovery, meaning we aimed to recover all the $k$ information bits with high probability. 
But depending on scenarios, it could be enough to recover only $\alpha k$ of information bits for $\alpha\in(0,1)$.
Then, the question is how much this relaxed recovery condition would help in reducing the sample complexity for a given query difficulty $\bard$.
Especially, one interesting question might be whether it is possible to recover $\alpha k$ information bits with only $n=\Theta(k)$ measurements even with the very low query difficulty $\bard=\Theta(1)$, which does not increase in $k$. In the exact recovery problem, it was impossible to reliably recover $k$ input symbols with the sample complexity $n=\Theta(k)$ when the query difficulty is $\bard=\Theta(1)$. With $\bard=\Theta(1)$, it was necessary to have at least $n=\Theta(k\log k)$ sample complexity for the exact recovery, which makes the ratio $k/n$ goes to 0 as $k\to\infty$. Therefore, it would be interesting to see whether the sample complexity of $n=\Theta(k)$ is sufficient for the partial recovery problem even with the query difficulty of $\bard=\Theta(1)$.

Another interesting direction is to apply the proposed query design to real crowdsourcing systems and to analyze the experimental trade-offs between the query difficulty and the sample complexity.
Especially, when the collected measurements contain inaccurate answers and the probability that the measurements include inaccurate answers changes depending on the query difficulty, the corresponding sample complexity might be a different function of $k$ and $\bard$. Therefore, it would be interesting to find the query difficulty that minimizes the sample complexity in crowdsourcing systems with random erasures and inaccurate answers, and this direction of research would help guiding the design of sample-efficient crowdsourcing systems.

\appendices
\section{Proof of Lemma~\ref{lem:comb}}\label{app:lem:comb}
To prove this lemma, we refer to the similar bound provided in~\cite{ahn2016community}.
\begin{lem}\label{lem:ahn}
{\it
Let $\beta=\ceil*{\max\left\{\frac{k-d+1}{2d+1},\frac{d+1}{2(k-d)+1}\right\}}$ and $\alpha=\max\left\{\frac{k-d+1}{d},\frac{d+1}{k-d}\right\}$. Then we have
\beq
\sum_{\substack{i\leq d\\ i \text{ is odd}}}{s\choose i}{k-s\choose d-i}\geq 
\begin{cases}
\frac{2s}{5\alpha}{k\choose d},& \text{when } s<\beta,\\
\frac{1}{5}{k\choose d}, & \text{when } \beta\leq s\leq k-\beta,\\
\frac{2(k-s)}{5\alpha}{k\choose d}, &\text{when }k-\beta<s.
\end{cases}
\eeq
}
\end{lem}

Note that
\beq
{k\choose d}=\sum_{\substack{i\leq d\\ i \text{ is odd}}}{s\choose i}{k-s\choose d-i}+\sum_{\substack{i\leq d\\ i \text{ is even}}}{s\choose i}{k-s\choose d-i}.
\eeq
Therefore, by using Lemma~\ref{lem:ahn}, we can find an upper bound on $\sum_{\substack{i\leq d\\ i \text{ is even}}}{s\choose i}{k-s\choose d-i}$ as a scaling of ${k \choose d}$ such that
 \beq
 \begin{aligned}
&I_d=\sum_{\substack{i\leq d\\ i \text{ is even}}}{s\choose i}{k-s\choose d-i}\\
&\leq
 \begin{cases} \label{eqn:I_d original bound}
\left(1- \frac{2s}{5\alpha}\right){k\choose d},& \text{when } s<\beta,\\
\frac{4}{5}{k\choose d}, & \text{when } \beta\leq s\leq k-\beta,\\
\left(1-\frac{2(k-s)}{5\alpha}\right){k\choose d}, &\text{when }k-\beta<s.
 \end{cases}
 \end{aligned}
 \eeq
We define
\beq
\kappa(s) = \frac{k-s+1}{2s+1}.
\eeq
We first consider the case $s \leq \frac{k}{2}$ (i.e., $s \leq k-s$). Since $\beta$ attains its maximum $\ceil*{\frac{k}{3}}$ at $d=1$ or $d=k-1$, we find that
$$
\beta < \frac{k}{2} \leq k-s.
$$
Hence, $k-\beta > s$ and the last case in \eqref{eqn:I_d original bound} cannot happen.

\begin{enumerate}
	\item For $d \leq \frac{k}{2}$ (or, $k-d \geq d$),
	$$
	\beta = \ceil*{\frac{k-d+1}{2d+1}}, \qquad \alpha = \frac{k-d+1}{d}.
	$$
	Note that $\frac{k-\kappa(s)+1}{2\kappa(s)+1} = s$. Since $\frac{k-d+1}{2d+1}$ is an increasing function of $d$, if $d < \kappa(s)$ then $\beta > s$. Thus,
	\beq
	I_d \leq \begin{cases}
	\left(1- \frac{2s}{5\alpha}\right){k\choose d},& \text{when } d < \kappa(s), \\
	\frac{4}{5}{k\choose d}, & \text{when } d \geq \kappa(s).
	\end{cases}
	\eeq
		
	\item For $d > \frac{k}{2}$ (or, $k-d < d$),
	$$
	\beta = \ceil*{\frac{d+1}{2(k-d)+1}}, \qquad \alpha = \frac{d+1}{k-d}.
	$$
	Proceeding as above, we get
	\beq
	I_d \leq \begin{cases}
	\left(1- \frac{2s}{5\alpha}\right){k\choose d},& \text{when } d > k-\kappa(s), \\
	\frac{4}{5}{k\choose d}, & \text{when } d \leq k-\kappa(s).
	\end{cases}
	\eeq

\end{enumerate}
In the case $s > \frac{k}{2}$, we can obtain the bounds for $I_d$ simply by changing $s$ to $k-s$. 

\section{Proof of Lemma~\ref{lem:kappas}}\label{app:lem:kappas}
In this lemma, we prove an upper bound on ${k\choose s}e^{-n\Sigma_s}$ where
\beq
\begin{split}
\Sigma_s &= \frac{1}{5} \sum_{d=\ceil*{\kappa(s)}}^{k-\ceil*{\kappa(s)}} \Omega_d + \\
&\frac{2s}{5} \left( \sum_{d=1}^{\ceil*{\kappa(s)}-1} \frac{d\,\Omega_d}{k-d+1} + \sum_{d=k-\ceil*{\kappa(s)}+1}^{k} \frac{(k-d)\Omega_d}{d+1} \right).
\end{split}
\eeq
For the Soliton distribution
\beq
\begin{aligned}\nonumber
\Omega_d = 
	\begin{cases}
	\frac{1}{D} & \text{ if } d=1 \\
	\frac{1}{d(d-1)} & \text{ if } 2\leq d \leq D \\
	0 & \text{ if } d > D\text{ or }d=0,
	\end{cases}
\end{aligned}
\eeq
we have the query difficulty
$$
\log (D+1) < \bard = \frac{1}{D} + \sum_{d=2}^{D} \frac{1}{d-1} = \sum_{d=1}^D \frac{1}{d} < \log D + 1.
$$
For simplicity, here we assume that $D \geq 2$.
Recall that
\beq
\kappa(s) = \frac{k-s+1}{2s+1},
\eeq
which is a decreasing function of $s$, and $\kappa(s) > 0$ for $s \leq \frac{k}{2}$.

\begin{enumerate}

\item If $\ceil*{\kappa(s)} > D$,
$$
\Sigma_s \geq \frac{2s}{5} \sum_{d=1}^{\ceil*{\kappa(s)}-1} \frac{d\,\Omega_d}{k-d+1} > \frac{2s}{5k} \sum_{d=1}^{D} d\,\Omega_d = \frac{2s \bar d}{5k}.
$$
Thus, if $n \bar d \geq 5 k \log k$,
$$
\binom{k}{s} e^{-n\Sigma_s} < k^s \exp \left( -\frac{2ns \bar d}{5k} \right) \leq k^s k^{-2s} = k^{-s}.
$$

\item If $4 \leq \ceil*{\kappa(s)} \leq D$, we first notice that
\beq \label{eq:s_kappa}
s < \frac{k-2}{7} \Leftrightarrow \kappa(s) > 3 \Leftrightarrow \ceil*{\kappa(s)} \geq 4.
\eeq
Thus $s \leq \frac{k-2}{7}$ and $\kappa(s) -1 = \frac{k-3s}{2s+1} \geq \frac{4k}{7(2s+1)} \geq \frac{4k}{21s}$. In this case,
\beq\nonumber
\begin{aligned}
\Sigma_s &\geq \frac{2s}{5} \sum_{d=1}^{\ceil*{\kappa(s)}-1} \frac{d\,\Omega_d}{k-d+1} \\
&> \frac{2s}{5k} \sum_{d=2}^{\ceil*{\kappa(s)}-1} \frac{1}{d-1}\\
& > \frac{2s}{5k} \log (\ceil*{\kappa(s)}-1)\\
& \geq \frac{2s}{5k} \log \left( \frac{4k}{21s} \right).
\end{aligned}
\eeq
Moreover, since $\frac{k}{s} \geq 7$, if $n \geq Ck$ for some sufficiently large $C$, ($C\geq 68$ suffices)
\beq\nonumber
\begin{aligned}
&n\Sigma_s \\
&\geq \frac{2Cs}{5} \log \left( \frac{4k}{21s} \right)\\
& \geq 4s \log \left( \frac{k}{s} \right) + \left( \frac{2C}{5} -4 \right) s \log 7 + \frac{2Cs}{5} \log \left( \frac{4}{21} \right)\\
& \geq 4s \log \left( \frac{k}{s} \right).
\end{aligned}
\eeq
From Stirling's formula, we also have that
$$
\sqrt{2\pi} n^{n+\frac{1}{2}} e^{-n} \leq n! \leq e n^{n+\frac{1}{2}} e^{-n},
$$
hence
\begin{equation*} \begin{split}
\binom{k}{s} 
&\leq \frac{e k^{k+\frac{1}{2}} e^{-k}}{2\pi (k-s)^{k-s+\frac{1}{2}} e^{-(k-s)} s^{s+\frac{1}{2}} e^{-s}} \\
&\leq \frac{\sqrt{k}}{2\sqrt{(k-s)s}} \cdot \frac{k^k}{(k-s)^{k-s} s^s} \\
&\leq  \left( \frac{k}{s} \right)^s \left(1-\frac{s}{k} \right)^{s-k} \\
&\leq \left( \frac{k}{s} \right)^s e^{s-\frac{s^2}{k}} = \exp \left( s \log \left( \frac{k}{s} \right) + s - \frac{s^2}{k} \right) \\
&\leq \exp \left( 2s \log \left( \frac{k}{s} \right) \right).
\end{split} \end{equation*}
Thus, if $n \geq 68k$,
$$
\binom{k}{s} e^{-n\Sigma_s} \leq \exp \left( -2s \log \left( \frac{k}{s} \right) \right) = \left( \frac{k}{s} \right)^{-2s}.
$$
Note that
$$
\left( \frac{k}{s} \right)^{-2s} \leq
	\begin{cases}
	k^{-s} & \text{ if } s \leq \sqrt k \,, \\
	2^{-2\sqrt k} &\text{ if } \sqrt k < s \leq k/2 \,.
	\end{cases}
$$
\item If $\ceil*{\kappa(s)} = 3$, we find from \eqref{eq:s_kappa} that $s \geq \frac{k-2}{7}$. Then, by considering the case $d=2$,
\beq\nonumber
\begin{aligned}
\Sigma_s &\geq \frac{2s}{5} \sum_{d=1}^{\ceil*{\kappa(s)}-1} \frac{d\,\Omega_d}{k-d+1}\\
& \geq \frac{2s}{5(k-1)}\\
& \geq \frac{2(k-2)}{35(k-1)} \\
&\geq \frac{1}{35}
\end{aligned}
\eeq
for $k \geq 3$. Thus, if $n \geq 35k$,
$$
\binom{k}{s} e^{-n\Sigma_s} \leq 2^k e^{-k}.
$$

\item If $\ceil*{\kappa(s)} = 1, 2$,
$$
\Sigma_s \geq \frac{1}{5} \sum_{d=\ceil*{\kappa(s)}}^{k-\ceil*{\kappa(s)}} \Omega_d \geq \frac{\Omega_2}{5} = \frac{1}{10}.
$$
Thus, if $n \geq 10k$,
$$
\binom{k}{s} e^{-n\Sigma_s} \leq 2^k e^{-k}.
$$
\end{enumerate}

\bibliographystyle{IEEEtran}

\end{document}